\documentclass{appolb}
\usepackage{graphicx}
\usepackage{lineno}
\usepackage{subcaption}
\usepackage{float}
\usepackage{tikz}
\usepackage{placeins}
\usepackage{hyperref}
\usepackage{amsmath,amssymb}

\begin{document}
\title{Recent results on Central Exclusive Production\\with the STAR detector at RHIC%
\thanks{Presented at Diffraction and Low-$x$ 2018 workshop, 26 August 2018 to 1 September 2018, Reggio Calabria, Italy, \href{https://indico.cern.ch/event/713101/}{https://indico.cern.ch/event/713101/}}%
}
\author{Rafa{\l} Sikora\thanks{This work was partially supported by the Polish National Science Centre grant No. UMO-2015/18/M/ST2/00162 and by the AGH UST grant No. 15.11.220.717/28 for 
young researchers within subsidy of Ministry of Science and Higher Education.} for the STAR Collaboration
\address{AGH University of Science and Technology,\\Faculty of Physics and Applied Computer Science,\\ al. A. Mickiewicza 30, 30-059 Krak\'ow, Poland}}
\maketitle
\begin{abstract}\vspace{-7pt}
We present preliminary differential fiducial cross sections on Central Exclusive Production (CEP, $pp\to pXp$) of two opposite-charge mesons 
($X=\pi^{+}\pi^{-}$, $K^{+}K^{-}$) in midrapidity region with small squared four-momentum transferred from forward protons, $0.03<|t_{1}|,|t_{2}|<0.2~\textrm{GeV}^{2}$. The process was measured in the STAR experiment at the Relativistic Heavy Ion Collider (RHIC) in proton-proton collisions at the center-of-mass energy $\sqrt{s}~=~200$~GeV.\vspace{-7pt}

\end{abstract}
  
\section{Introduction}

The STAR experiment at RHIC in recent years puts a lot of effort into the measurements of high-energy particle diffraction processes~\cite{wlodek,lukasz}, such as Central Exclusive Production (CEP) in proton-proton collisions. In this process, the protons stay intact after the interaction, while the fraction part of their momentum is transferred into the neutral system measured by the central detector. The transverse momenta of the protons gained from this scattering process makes them deviate from the nominal beamline trajectory which enables their detection in forward detectors placed close to the beamline.

The dominant CEP mechanism expected at studied center-of-mass energy is the Double Pomeron Exchange (DPE) - a fusion of two Pomerons in the Regge framework. 
The gluon-rich environment of the reaction is considered as suitable for the search of the gluon bound states (glueballs), which are predicted by the QCD theory but not yet confirmed.

\section{Experimental setup}
Thanks to the wide acceptance of installed detectors, measurement of CEP in the STAR experiment is possible. The heart of the~STAR detector, the Time Projection Chamber (TPC), is used to reconstruct momentum and specific energy loss ($dE/dx$) of charged particles in pseudorapidity range $|\eta|<1$. The Time-Of-Flight (TOF) system surrounding TPC is used to extend particle identification capabilities by measuring particle velocity.
Last but not least, there are Roman Pot (RP) detectors which allow efficient triggering and measurement of diffractively scattered protons. Detailed description of STAR forward proton detectors can be found in Ref.~\cite{cep_star}.\vspace*{-5pt}

\section{Event selection}
The presented preliminary results were obtained using dedicated data collected in year 2015 during 11-week period of $p+p$ collisions at $\sqrt{s}=200$~GeV. 
The trigger logic required signals in RP scintillators on both sides of the STAR central barrel to ensure the detection of scattered protons. In addition, at least 2 trigger-level hits in the TOF detector were required to ensure presence of at least 2 charged particles in the TPC. Additionally, a veto was performed on signal in small Beam-Beam Counter (BBC) scintillator tiles ($3.3<|\eta|<5.0$) and Zero-Degree Calorimeters (ZDCs) to assure no overlap with pile-up interactions which would fill the rapidity gaps expected in DPE between final state protons and central diffractive system.

\begin{figure}[b]\vspace*{-10pt}
	\parbox{0.5\textwidth}{
		\caption[]{Distribution (raw counts) of the $x$- (horizontal axis) and $y$-component (vertical axis) of the forward proton momentum in events passing all CEP event selection cuts, except forward proton fiducial region requirement which is drawn with solid black lines. Parts of the distribution at positive and negative $p_{y}$ correspond to protons measured in the detectors placed above and below the beamline, respectively.}\label{fig:pxpy}
	}
	\parbox{0.5\textwidth}{
		\centering%
		\includegraphics[width=\linewidth]{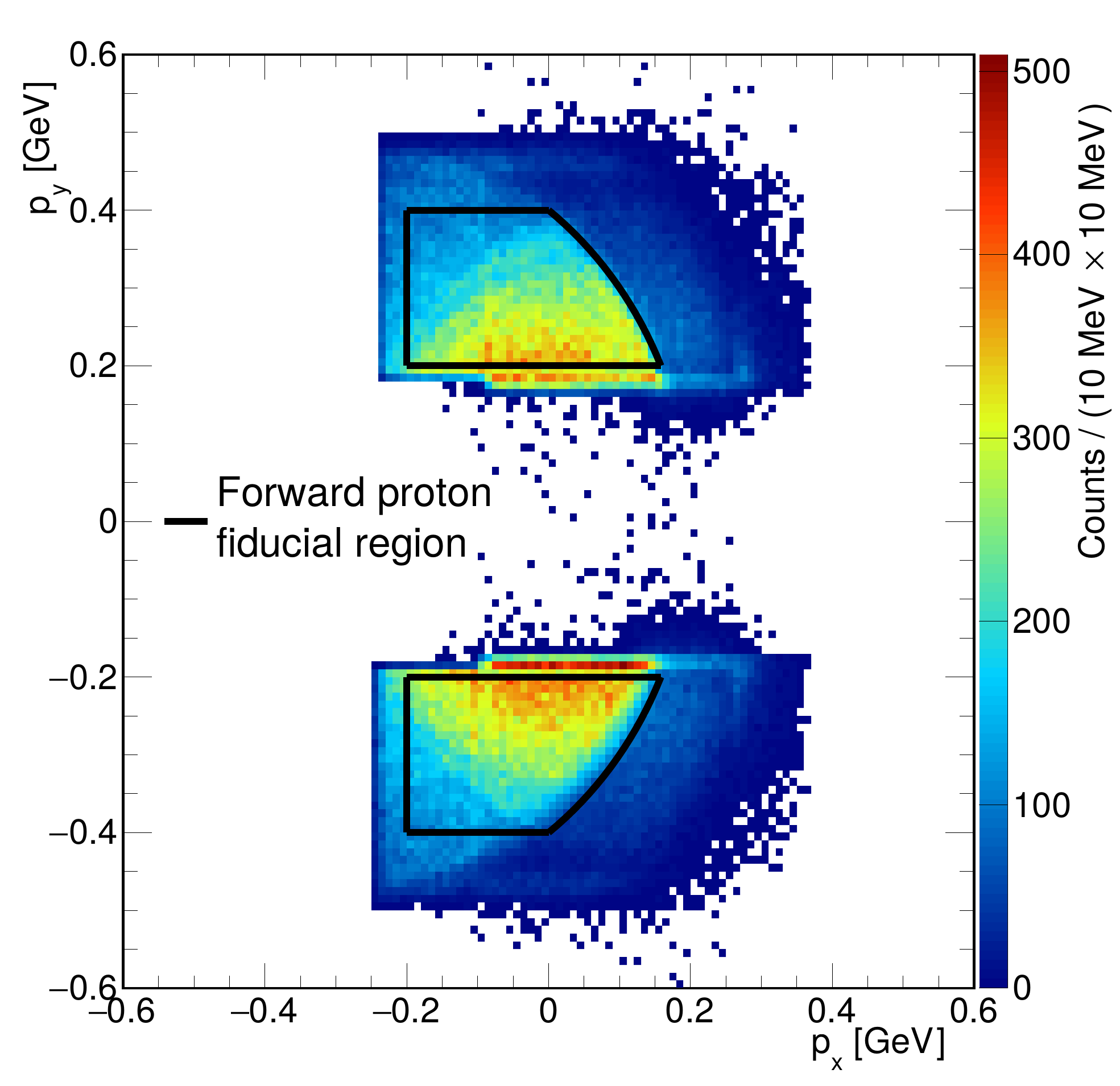}
	}
\end{figure}%

Analysis of CEP events started with selection of good quality proton tracks reconstructed in RPs on both sides of the STAR detector. These protons were further required to have transverse momentum ($p_{x}$, $p_{y}$) in the fiducial region that guaranteed high geometrical acceptance, and overall trigger, and track reconstruction efficiency (Fig.~\ref{fig:pxpy}). The fiducial region 
was chosen based on detailed simulations of diffractive proton propagation from the interaction point through collider lattice (beampipe, magnets) to RP detectors, implemented in Geant4.

\begin{figure}[b]\vspace*{-10pt}
	\parbox{0.5\textwidth}{
		\centering
		\begin{subfigure}[b]{0.975\linewidth}{\label{fig:pTmiss}
				{\includegraphics[width=\linewidth]{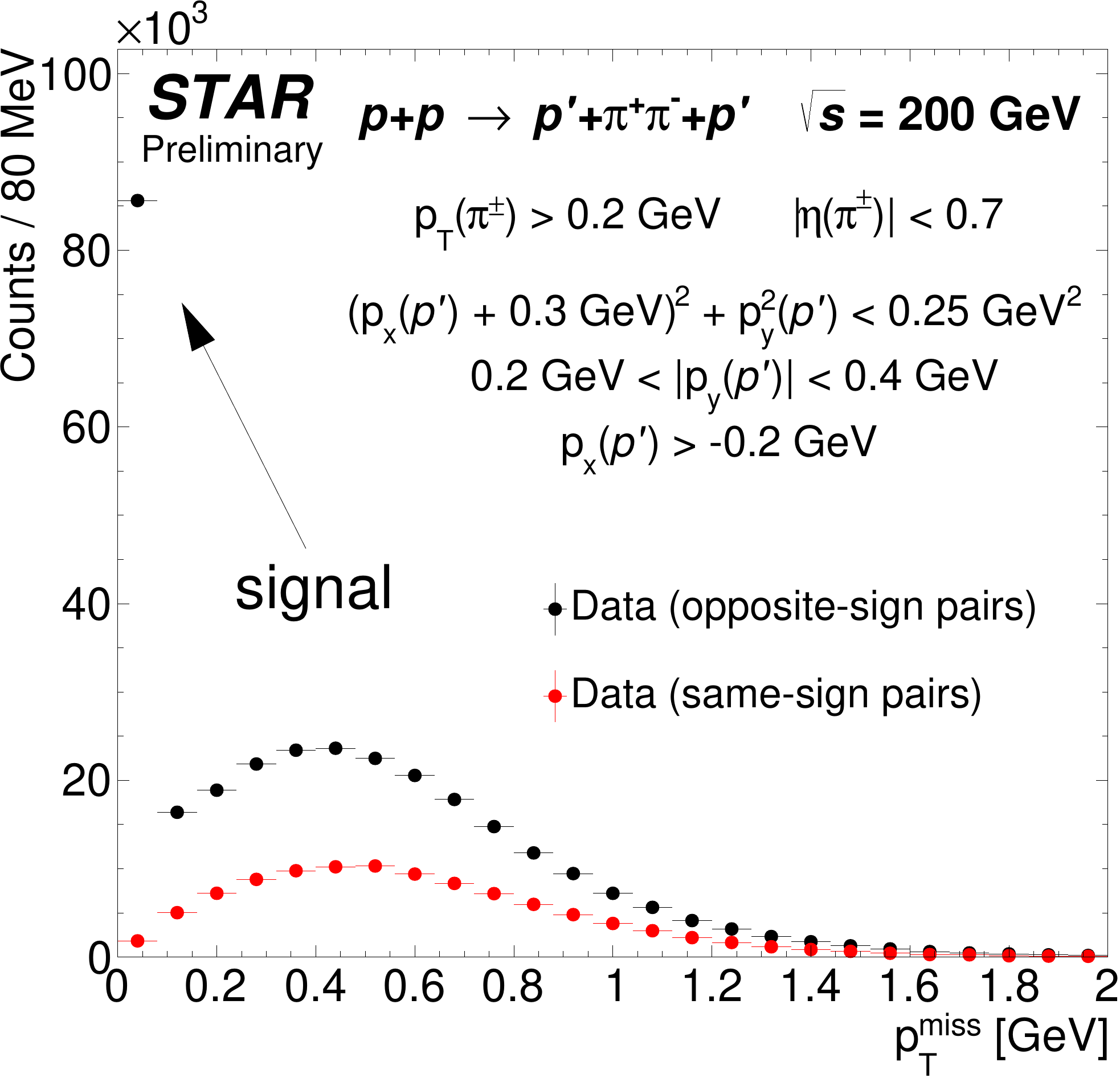}}}
		\end{subfigure}
	}
	\parbox{0.5\textwidth}{
		\centering
		\begin{subfigure}[b]{0.975\linewidth}{\label{fig:invMassRaw}\vspace{-2pt}
				{\includegraphics[width=0.98\linewidth]{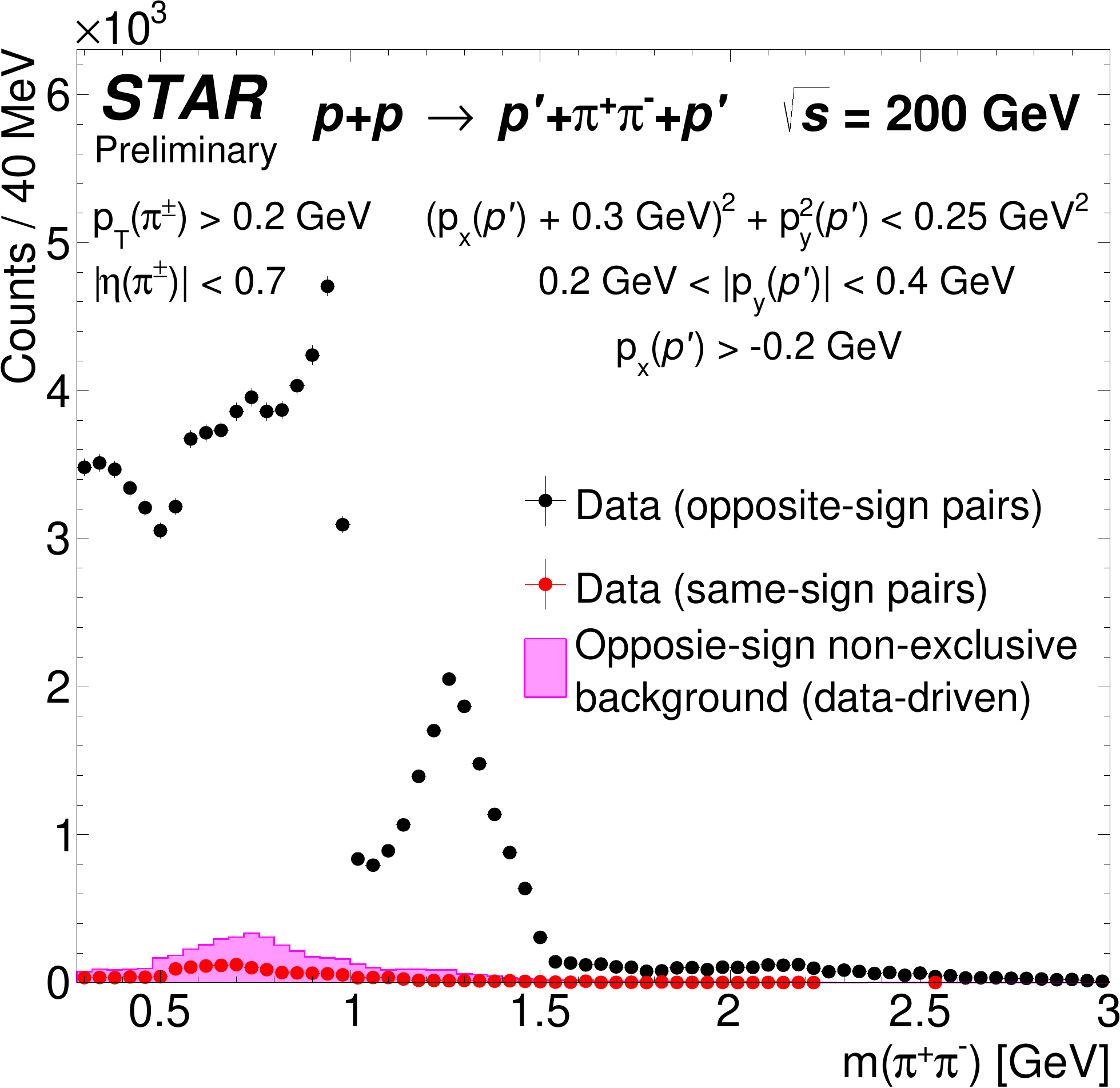}}}
		\end{subfigure}
	}
	\caption[]{(left) Distribution (raw counts) of the total transverse momentum and (right) invariant mass of $\pi^{+}\pi^{-}$ in the fiducial region defined in the plots. The same-sign background events are also shown. Filled violet histogram in the invariant mass plot represents the opposite-sign background determined from the data using $p_{\text{T}}^{\text{miss}}$ extrapolation from the background-only region ($p_{\text{T}}^{\text{miss}}>200~\text{MeV}$) to 0 in bins of $m(\pi^{+}\pi^{-})$.}
	\label{fig:pTmiss_invMassRaw}
\end{figure}

Next, exactly 2 opposite-sign TPC tracks matched with hits in TOF were required, both originating from the only primary vertex in an event and passing track quality cuts with $p_{\text{T}}>0.2$~GeV and $|\eta|<0.7$. The $z$-position of the vertex was limited to $|z_{\text{vtx}}|<80$~cm to ensure high geometrical acceptance for the central tracks in the entire fiducial phase space. The $z$-position of the vertex could also be reconstructed by measuring the time difference of the two protons reconstructed from RPs (resolution $\sigma_{\Delta z_{\text{vtx}}}=$~11~cm), which was compared to that measured in TPC -  difference of $3\sigma_{\Delta z_{\text{vtx}}}$ was allowed.

To further suppress the residual backgrounds, a veto on minimum ionizing particle signal in large BBC tiles ($2.1<|\eta|<3.3$) was imposed, as well as veto on additional TOF hits without matching TPC tracks. This mainly removed higher multiplicity central diffraction events with some particles not reconstructed or produced outside of TPC and TOF acceptance.

Identification of particles was performed using combined information from TPC and TOF. A pair of tracks was recognized as either $\pi^{+}\pi^{-}$, $K^{+}K^{-}$ or $p\bar{p}$ if the $dE/dx$ of both tracks was consistent with given hypothesis and mass of particle calculated from the time of hit in TOF (assuming equal masses of positive and negative charge particle) also supported that hypothesis. If a pair was recognized as species other than $\pi^{+}\pi^{-}$, an additional cut on track $p_{\text{T}}$ was imposed: $>0.3$~GeV for $K^{\pm}$ and $>0.4$~GeV for $p(\bar{p})$.

After the event selection described above, one can clearly see the signal events among remaining background in the distribution of total transverse momentum of all measured particles, $p_{\text{T}}^{\text{miss}}$ (Fig.~\ref{fig:pTmiss_invMassRaw}, left). A final cut on total transverse momentum, $p_{\text{T}}^{\text{miss}}<75$~MeV, was employed to improve the exclusivity of the system. The cut value was determined by the finite resolution of transverse momentum originated from the angular divergence of proton beams.

Fig.~\ref{fig:pTmiss_invMassRaw} (right) shows raw distribution of the invariant mass of $\pi^{+}\pi^{-}$ system in CEP  process. One can see that the background
does not exceed a few percent level. Overall numbers of selected $\pi^{+}\pi^{-}$, $K^{+}K^{-}$ and $p\bar{p}$ pairs equals 84~000, 1~000 and 70, respectively.\vspace*{-9pt}

\section{Results}
\begin{figure}[t]
	\parbox{0.5\textwidth}{
		
		\begin{subfigure}[b]{0.975\linewidth}{\label{fig:invMassPiPi}
				{\includegraphics[width=\linewidth]{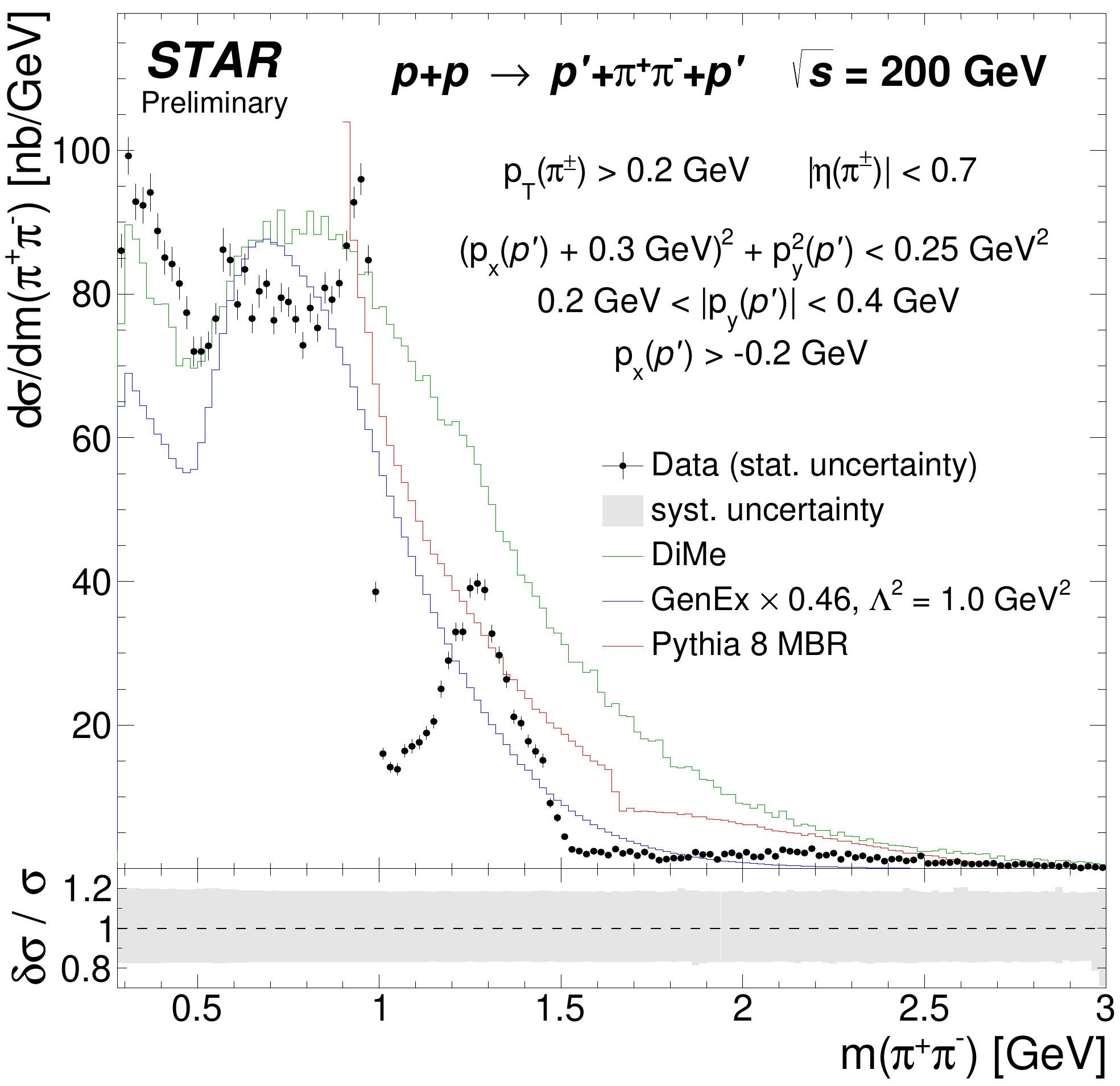}}}
		\end{subfigure}
	}
	\parbox{0.5\textwidth}{
		
		\begin{subfigure}[b]{0.975\linewidth}{\label{fig:invMassKK}
				{\includegraphics[width=\linewidth]{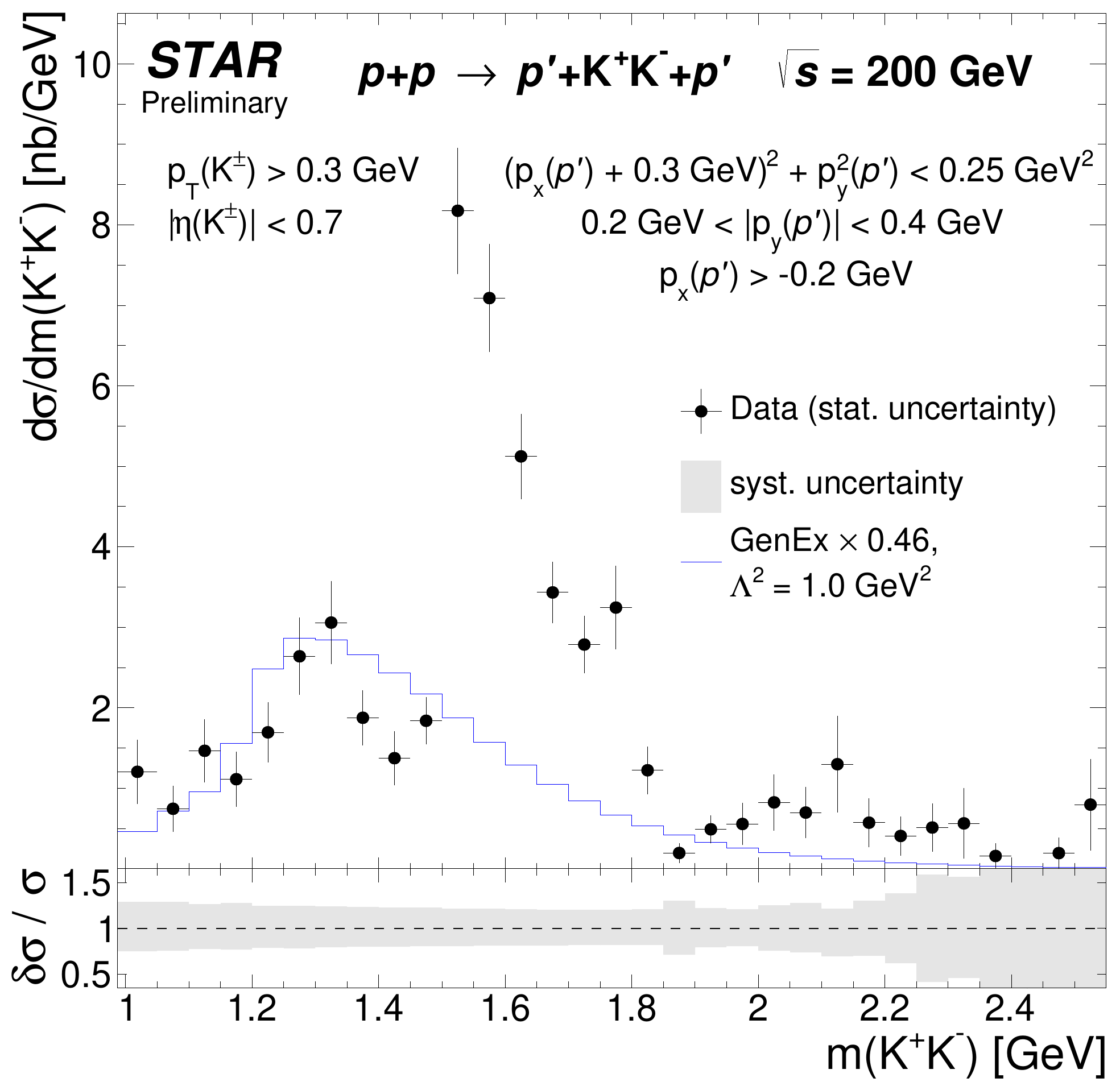}}}
		\end{subfigure}
	}
	\caption[]{(left) Differential fiducial cross section as a function of invariant mass of $\pi^{+}\pi^{-}$ and (right) invariant mass of $K^{+}K^{-}$ compared with some models of non-resonant CEP.}
	\label{fig:invMassPiPi_invMassKK}\vspace*{-15pt}
\end{figure}

\begin{figure}[b]\vspace*{-5pt}
	\parbox{0.5\textwidth}{
		\centering
		\begin{subfigure}[b]{0.975\linewidth}{\label{fig:deltaPhi}
				{\includegraphics[width=\linewidth]{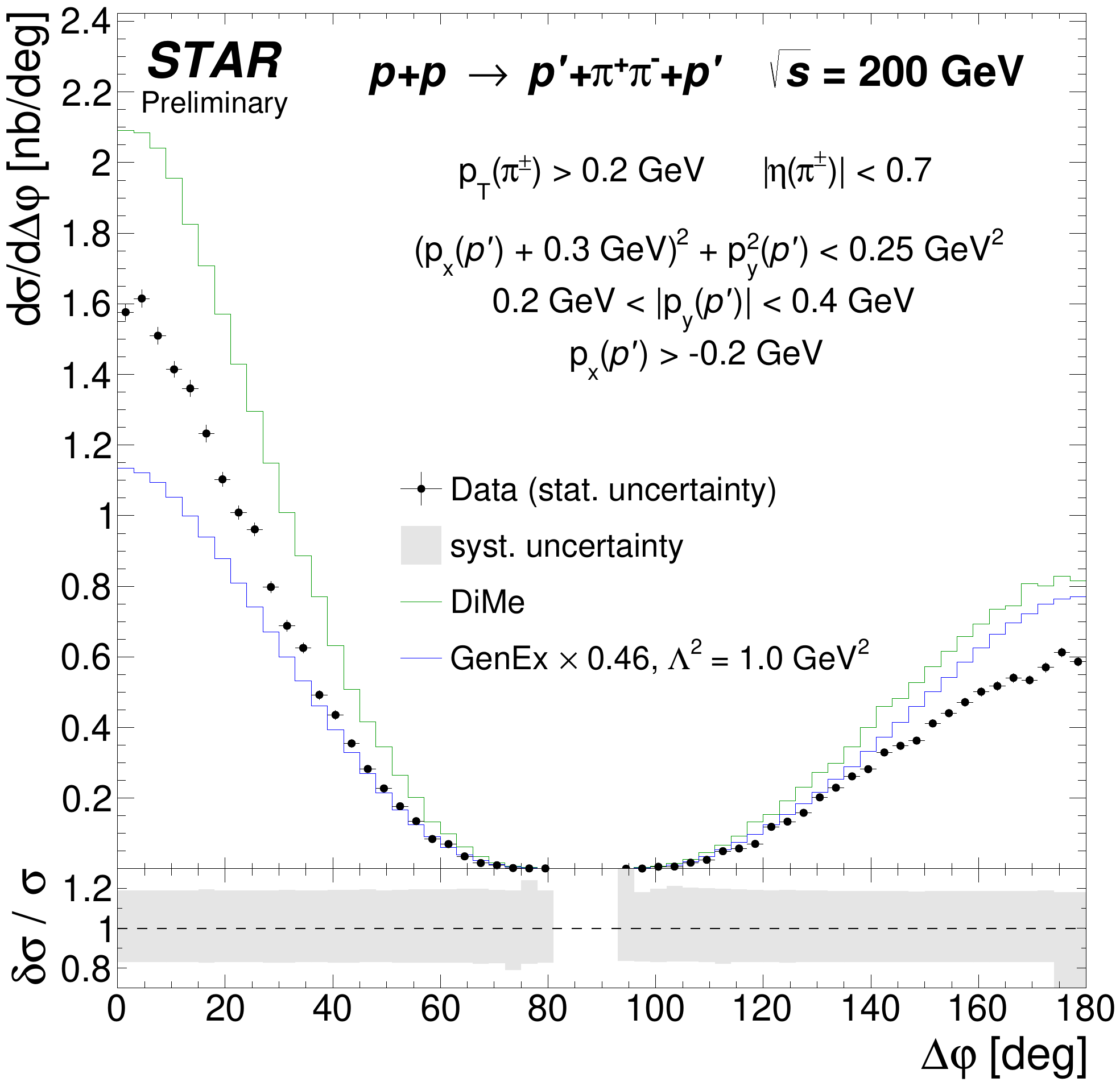}}}
		\end{subfigure}
	}
	\parbox{0.5\textwidth}{
		\centering
		\begin{subfigure}[b]{0.975\linewidth}{\label{fig:InvMassVsDeltaPhi}
				{\includegraphics[width=\linewidth]{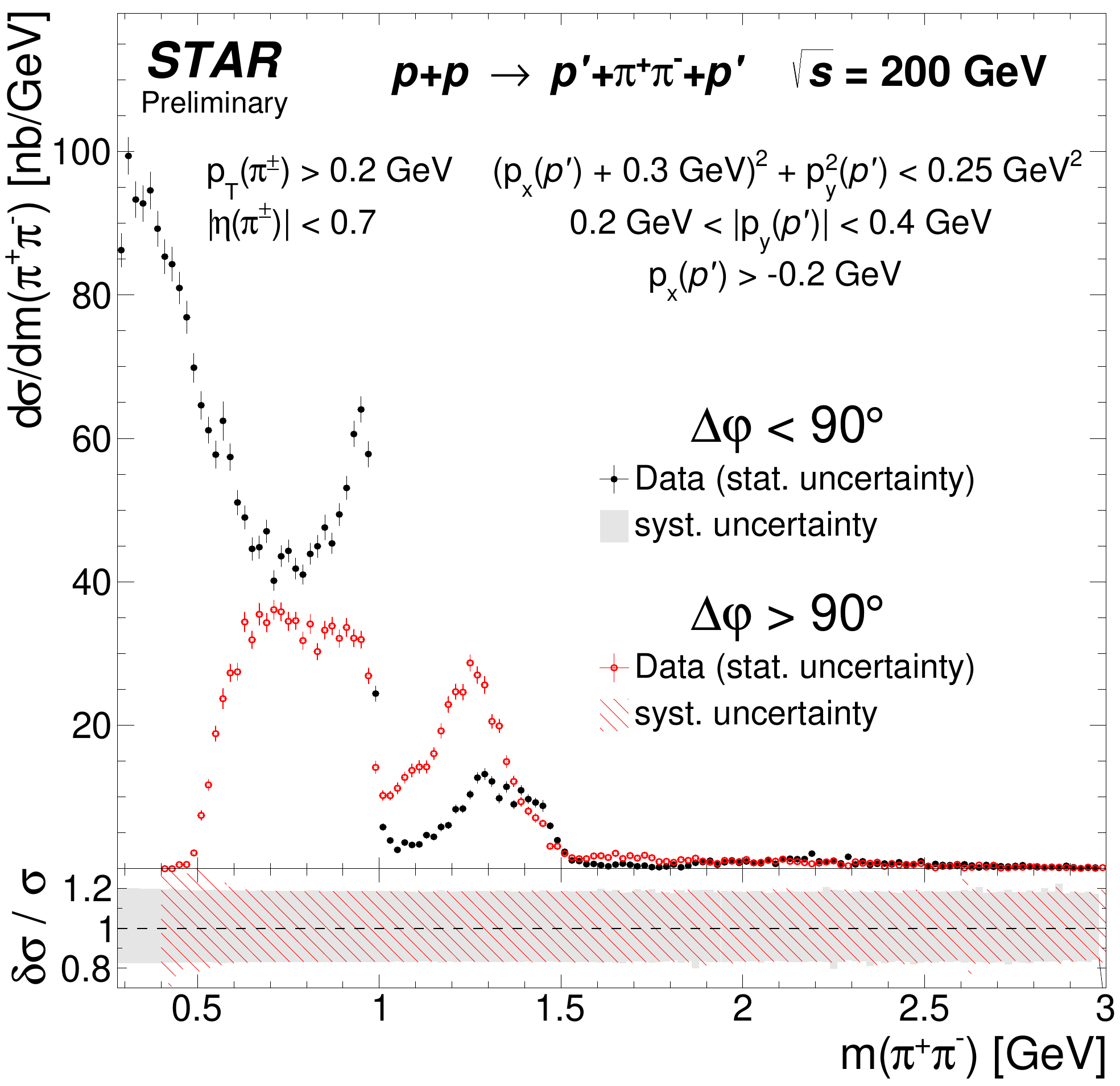}}}
		\end{subfigure}
	}
	\caption[]{(left) Differential fiducial cross section as a function of the angular separation between scattered forward protons in the transverse plane $\Delta\varphi$ and (right) invariant mass of $\pi^{+}\pi^{-}$ for two ranges of $\Delta\varphi$.}
	\label{fig:deltaPhi_InvMassVsDeltaPhi}\vspace*{-15pt}
\end{figure}

\begin{figure}
	\parbox{0.5\textwidth}{
		\centering
		\begin{subfigure}[b]{0.975\linewidth}{\label{fig:pairRapidity}
				{\includegraphics[width=\linewidth]{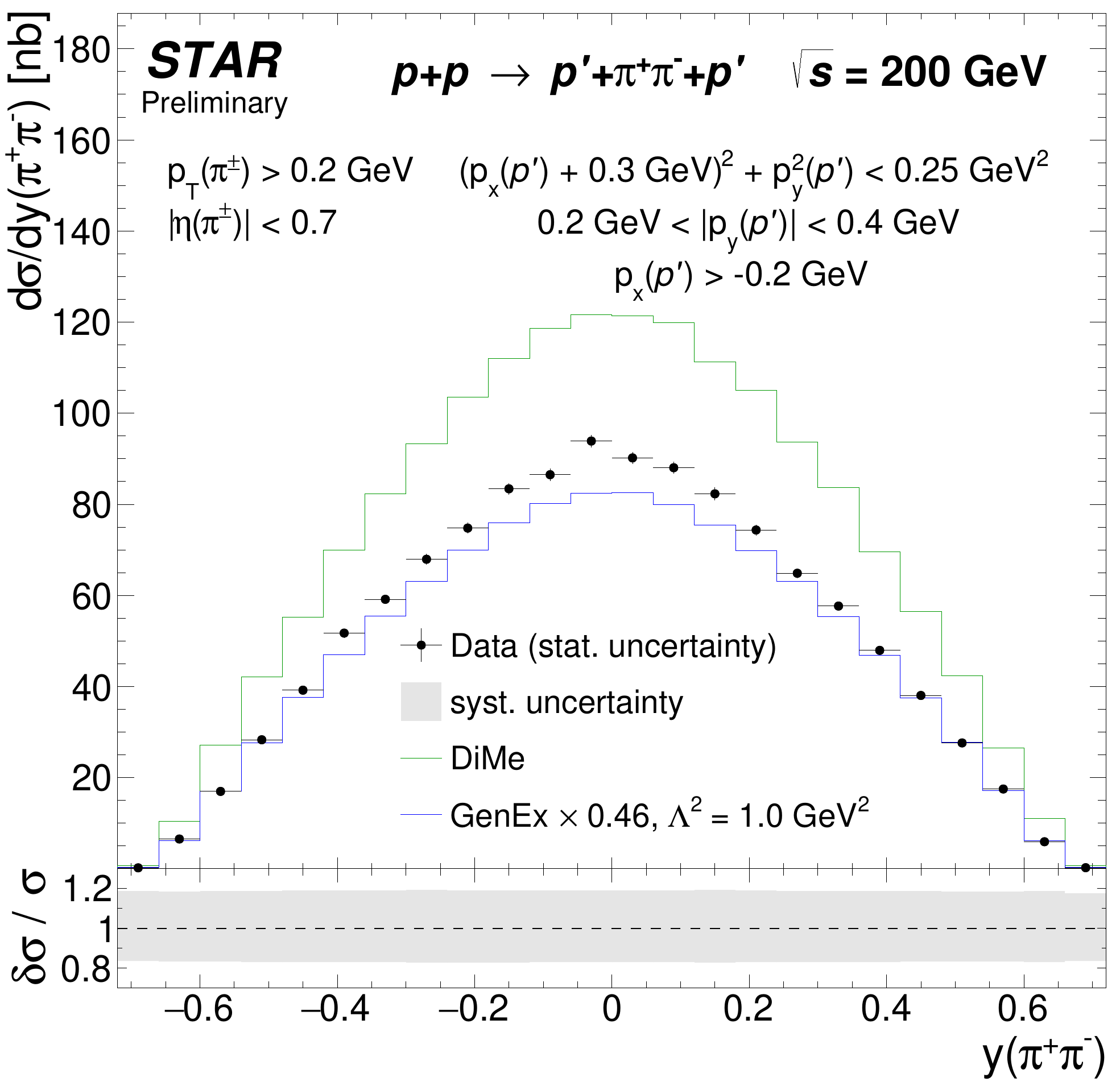}}}
		\end{subfigure}
	}
	\parbox{0.5\textwidth}{
		\centering
		\begin{subfigure}[b]{0.975\linewidth}{\label{fig:MandelstamTSum}
				{\includegraphics[width=\linewidth]{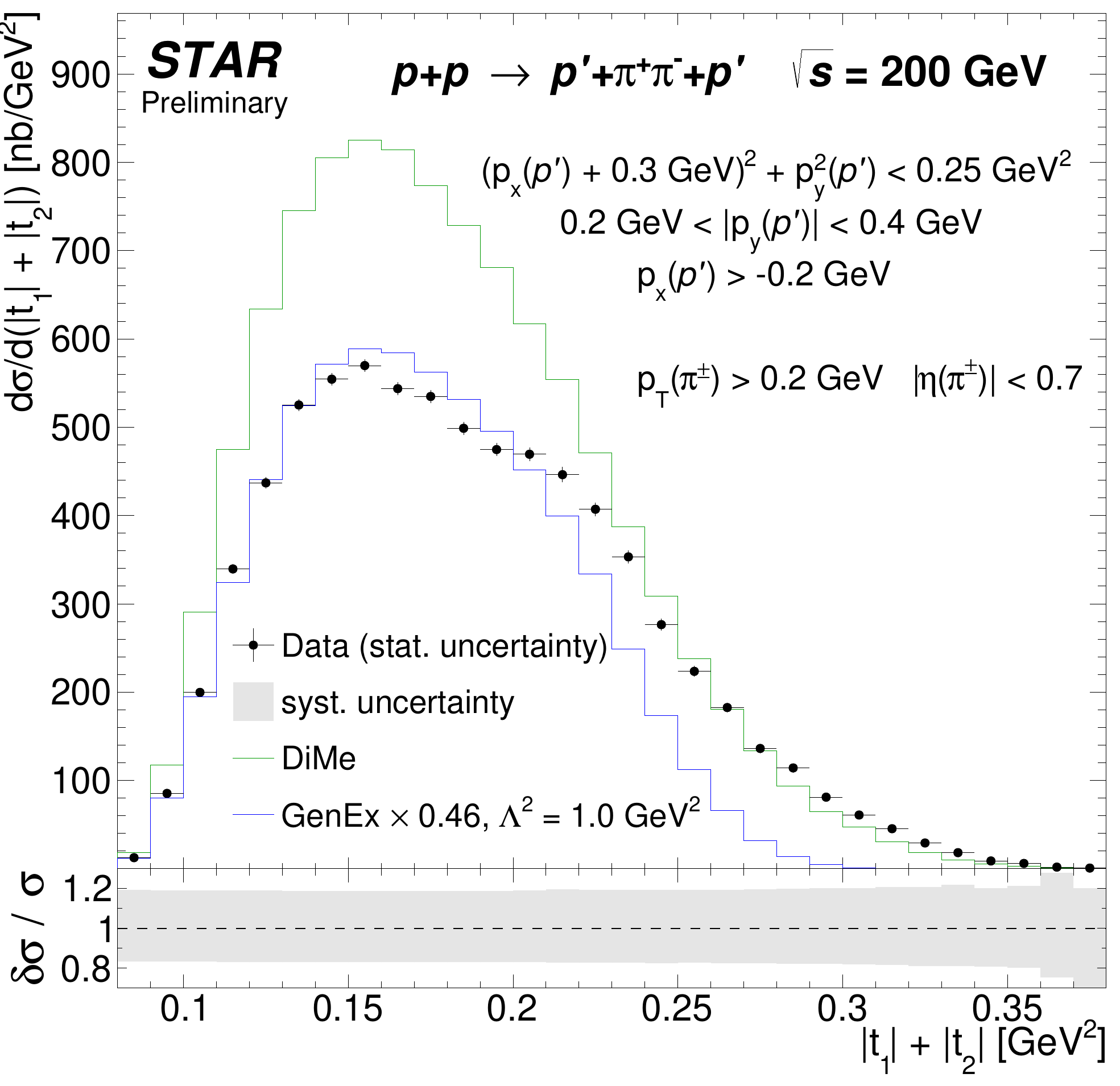}}}
		\end{subfigure}
	}
	
	\caption{(left) Differential fiducial cross section as a function of the rapidity of $\pi^{+}\pi^{-}$ pair and (right) sum of squared four-momentum carried by objects exchanged by protons in CEP of $\pi^{+}\pi^{-}$.}
	\label{fig:pairRapidity_MandelstamTSum}\vspace*{-10pt}
\end{figure}

Events selected with the cut flow described above were subjected to background subtraction and model-independent efficiency corrections derived from Monte Carlo (MC) simulations of the STAR detector response to particles traversing the central detector and protons passing through RP detectors. A technique of full MC event embedding into zero-bias data was utilized to reproduce all real environment effects. Closure tests were also conducted on MC events to verify that distributions of reconstructed and corrected observables are consistent with those at the hadron level. Finally, the differential fiducial cross sections were calculated, presented in Figs.~\ref{fig:invMassPiPi_invMassKK}-\ref{fig:pairRapidity_MandelstamTSum}. For each result a conservative systematic uncertainty was estimated and drawn with the gray band in the panel below main plot. Systematic uncertainties are strongly correlated between data points. Luminosity uncertainty (10\%) was also included.

The invariant mass spectra of exclusively produced $\pi^{+}\pi^{-}$ and $K^{+}K^{-}$ pairs are presented in Fig.~\ref{fig:invMassPiPi_invMassKK}, together with the predictions of some non-resonant pair production models implemented in MC event generators~\cite{dime,genex,pythia}. Models qualitatively describe $d\sigma/dm(\pi^{+}\pi^{-})$ up to $\sim0.7$~GeV, which is consistent with the expectation of no narrow resonances in this mass range. The peak in data below 1 GeV followed by a sharp drop of the cross section is presumably due to $f_{0}(980)$ interference with production amplitude opposite in phase to $\pi^{+}\pi^{-}$ continuum. The peak between 1-1.5~GeV probably originated from $f_{2}(1270)$ with potential mixture of $f_{0}(1370)$, while the sharp drop of the cross section at 1.5~GeV might be due to $f_{0}(1500)$ interference with the remaining states.

In the $K^{+}K^{-}$ mass spectrum, the key feature is the prominent peak starting at $1.5~\textrm{GeV}$ which could be the $f_{0}(1500)$ state mixed with $f'_{2}(1525)$ and $f_{0}(1710)$. There is also a structure in the $f_{2}(1270)$ mass region, however this part could be well described by the non-resonant model only.

In Fig.~\ref{fig:deltaPhi_InvMassVsDeltaPhi} (left) we show differential fiducial cross section for CEP of $\pi^{+}\pi^{-}$ as a function of the azimuthal angle between scattered protons, $\Delta\varphi$, which is equal to the incident angle between objects exchanged by protons in the $xy$-plane. In Fig.~\ref{fig:deltaPhi_InvMassVsDeltaPhi} (right) we present $d\sigma/dm(\pi^{+}\pi^{-})$ in two ranges of $\Delta\varphi$. One can clearly notice significant differences between these subsets. The low cross section below $\sim$0.7~GeV for $\Delta\varphi>90^{\circ}$ is a kinematic effect connected with the low transverse momentum of $\pi^{+}\pi^{-}$ pair (pions do not reach requested $p_{\text{T}}>$~0.2~GeV). However, above that mass the difference in shape is driven only by the dependence of production amplitudes of various states on dynamics of DPE.

In Fig.~\ref{fig:pairRapidity_MandelstamTSum} we introduce differential fiducial cross section as a function of the central pair rapidity and sum of Mandelstam's $t$ variables. The data can not be described by continuum models which confirms large contribution from resonances in CEP. The differences may also indicate significant role of absorption effects which influence the cross sections.\vspace*{-4pt}

\section{Summary}
The preliminary STAR results on CEP of $\pi^{+}\pi^{-}$ and $K^{+}K^{-}$ in proton-proton collisions at $\sqrt{s}=$~200~GeV with detected forward scattered protons have been presented and discussed. In general, models of non-resonant CEP can not describe the data, which agrees with expectation of significant role of resonance production.\vspace{-20pt}

\end{document}